# Further Study on the Conservation Laws of Energy-momentum Tensor Density for a Gravitational System

Chen Fang-Pei

Department of Physics, Dalian University of Technology, Dalian 116024, China.

E-mail: chenfap@dlut.edu.cn

**Abstract** The various methods to derive Einstein conservation laws and the relevant definitions of energy-momentum tensor density for gravitational fields are studied in greater detail. It is shown that these methods are all equivalent. The study on the identical and different characteristics between Lorentz and Levi-Civita conservation laws and Einstein conservation laws is thoroughly explored. Whether gravitational waves carry the energy-momentum is discussed and some new interpretations for the energy exchanges in the gravitational systems are given. The viewpoint that PSR1913 does not verify the gravitational radiation is confirmed.

## 1. Introduction

There exist the following two kinds of conservation laws of energy-momentum tensor density for a gravitational system [1, 2, 3]:

I, Lorentz and Levi-Civita conservation laws

$$\frac{\partial}{\partial x^\mu}(\sqrt{-g}T^\mu_{(M)\alpha} + \sqrt{-g}T^\mu_{(G)\alpha}) = 0 \tag{1}$$

$$\sqrt{-g}T^\mu_{(M)\alpha} + \sqrt{-g}T^\mu_{(G)\alpha} = 0 \tag{2}$$

II, Einstein conservation laws

$$\frac{\partial}{\partial x^\mu}(\sqrt{-g}T^\mu_{(M)\alpha} + \sqrt{-g}\tilde{t}^\mu_{(G)\alpha}) = 0 \tag{3}$$

In Eq.(1), $\sqrt{-g}T^\mu_{(M)\alpha}$ is the energy-momentum tensor density for matter field, $\sqrt{-g}T^\mu_{(G)\alpha}$ is the energy-momentum tensor density for gravitational field; and $\sqrt{-g}T^\mu_{(G)\alpha}$ relates to $\sqrt{-g}T^\mu_{(M)\alpha}$ by Eq.(2).

In Eq.(3), $\sqrt{-g}T^{\mu}_{(M)\alpha}$ is also the energy-momentum tensor density for matter field, which has the same definition as in Eq.(1); $\sqrt{-g}\tilde{t}^{\mu}_{(G)\alpha}$ is the energy-momentum pseudo tensor density for gravitational field, which is not tensor density! The relationship between $\sqrt{-g}\tilde{t}^{\mu}_{(G)\alpha}$ of Eq.(3) and $\sqrt{-g}T^{\mu}_{(G)\alpha}$ of Eq.(1) can be determined by the Lagrangian density of gravitational field $\sqrt{-g(x)}L_G(x)$. For example, if the Lagrangian density of gravitational field is the generalized Einstein's Lagrangian density

$$\sqrt{-g(x)}L_G(x) = \sqrt{-g(x)}L_G[g_{\mu\nu}(x); g_{\mu\nu,\lambda}(x); g_{\mu\nu,\lambda\sigma}(x)] \tag{4}$$

(where the metric tensor $g_{\mu\nu}$ is used as the dynamical gravitational field), there exists the relation[2]:

$$\sqrt{-g}\tilde{t}^{\mu}_{(G)\alpha} = \sqrt{-g}T^{\mu}_{(G)\alpha} - \frac{\partial}{\partial x^{\beta}}u^{\mu\beta}_{(G)\alpha}, \quad (\frac{\partial}{\partial x^{\beta}}u^{\mu\beta}_{(G)\alpha} = -\frac{\partial}{\partial x^{\beta}}u^{\beta\mu}_{(G)\alpha}) \tag{5}$$

Where $u^{\lambda\mu}_{(G)\alpha} = \frac{\partial}{\partial x^{\sigma}}(\frac{\partial(\sqrt{-g}L_G)}{\partial g_{\nu\lambda,\mu\sigma}}g_{\nu\alpha})$. If the Lagrangian density of gravitational field $\sqrt{-g(x)}L_G(x)$ is different from Eq.(4), the definition of $u^{\lambda\mu}_{(G)\alpha}$ is different from that in Eq.(5)[1,3].

Eq.(4) is a simple Lagrangian density of gravitational field which is close to the original formulation of general relativity. For the sake of simplicity, we shall use Eq (4) as the Lagrangian density of gravitational field in this paper. For more complicated cases, please refer to Ref.[1,3].

Many general characters and peculiarities of Eq.(1) and Eq.(3) have been described and discussed in Ref.[1,2,3]. In this paper we shall delve into some other characters and peculiarities of Eq.(1) and Eq.(3) which was not discussed in Ref.[1,2,3] yet. The main topics to be discussed are:

(1) It will be shown that the various methods to derive Einstein conservation laws are equivalent, and the identical and different characteristics between Lorentz and Levi-Civita conservation laws and Einstein conservation laws will be explored thoroughly.

(2) Whether the gravitational waves carry the energy-momentum is discussed, and some new interpretations for the energy exchanges in the gravitational systems are given. The viewpoint that PSR1913 does not verify the gravitational radiation is confirmed.

**2. The equivalence of various methods to derive Einstein conservation laws**

In this section we shall let the Lagrangian density of matter field and gravitational field be

$\sqrt{-g(x)} L_M(x) = \sqrt{-g(x)} L_M [\psi(x); \psi_{,\lambda}(x); g_{\mu\nu}(x); g_{\mu\nu,\lambda}(x)]$ and

$\sqrt{-g(x)} L_G(x) = \sqrt{-g(x)} L_G [g_{\mu\nu}(x); g_{\mu\nu,\lambda}(x); g_{\mu\nu,\lambda\sigma}(x)]$ respectively, and use Einstein field equations $\sqrt{-g}(R^{\mu\nu} - \frac{1}{2} g^{\mu\nu} R) = -8\pi G \sqrt{-g} T^{\mu\nu}_{(M)}$ as the gravitational field equations. The energy-momentum tensor density of matter field is always defined as[2,4,5]

$$\sqrt{-g} T^{\mu}_{(M)\alpha} \stackrel{def}{=} 2 \frac{\delta(\sqrt{-g} L_M)}{\delta g_{\mu\nu}} g_{\nu\alpha} = 2\left( \frac{\partial(\sqrt{-g} L_M)}{\partial g_{\mu\nu}} - \frac{\partial}{\partial x^\lambda} \frac{\partial(\sqrt{-g} L_M)}{\partial g_{\mu\nu,\lambda}} \right) g_{\nu\alpha} \quad (6)$$

But there are different definitions of energy-momentum tensor density for gravitational field. For Lorentz and Levi-Civita conservation laws, the energy-momentum tensor density of gravitational field is defined as[1,2,3]

$$\sqrt{-g} T^{\mu}_{(G)\alpha} = 2 \frac{\delta(\sqrt{-g} L_G)}{\delta g_{\mu\nu}} g_{\nu\alpha} = 2\left( \frac{\partial(\sqrt{-g} L_G)}{\partial g_{\mu\nu}} - \frac{\partial}{\partial x^\lambda} \frac{\partial(\sqrt{-g} L_G)}{\partial g_{\mu\nu,\lambda}} + \frac{\partial^2}{\partial x^\lambda \partial x^\sigma} \frac{\partial(\sqrt{-g} L_G)}{\partial g_{\mu\nu,\lambda\sigma}} \right) g_{\nu\alpha}$$

(7)

Owing to $\frac{\delta(\sqrt{-g} L_M + \sqrt{-g} L_G)}{\delta g_{\mu\nu}} = 0$, there must have $\sqrt{-g} T^{\mu}_{(M)\alpha} + \sqrt{-g} T^{\mu}_{(G)\alpha} = 0$.

As for Einstein conservation laws, there are various methods to derive it. In Ref.[1,2] we use the symmetry of the Lagrangian density to obtain the conservation laws

$\frac{\partial}{\partial x^\mu} ( \sqrt{-g} T^{\mu}_{(M)\alpha} + \sqrt{-g} \tilde{t}^{\mu}_{(G)\alpha} ) = 0$; Where $\sqrt{-g} \tilde{t}^{\mu}_{(G)\alpha}$ is defined by

$$\sqrt{-g} \tilde{t}^{\mu}_{(G)\alpha} \stackrel{def}{=} \sqrt{-g} L_G \delta^{\mu}_{\alpha} - \frac{\partial(\sqrt{-g} L_G)}{\partial g_{\rho\nu,\mu}} g_{\rho\nu,\alpha} - \frac{\partial(\sqrt{-g} L_G)}{\partial g_{\rho\nu,\mu\sigma}} g_{\rho\nu,\sigma\alpha} + \frac{\partial}{\partial x^\sigma} \left( \frac{\partial(\sqrt{-g} L_G)}{\partial g_{\rho\nu,\mu\sigma}} \right) g_{\rho\nu,\alpha}$$

(8)

and there are the relations

$$\sqrt{-g}\tilde{t}^{\mu}_{(G)\alpha} = \sqrt{-g}T^{\mu}_{(G)\alpha} - \frac{\partial}{\partial x^{\beta}}u^{\mu\beta}_{(G)\alpha} \quad (\frac{\partial}{\partial x^{\beta}}u^{\mu\beta}_{(G)\alpha} = -\frac{\partial}{\partial x^{\beta}}u^{\beta\mu}_{(G)\alpha})_{\circ}$$

In some gravitational theory text books [4,5,6], the writers use Einstein field equations and/or Bianchi identities to derive Einstein conservation laws. There are three different methods listed below:

**A. The method used in the text book "The theory of relativity" [6] written by Moller**

From Bianchi identities, we have [5,6,7]

$$(\sqrt{-g}T^{\mu}_{(M)\alpha})_{;\mu} = \frac{\partial}{\partial x^{\mu}}(\sqrt{-g}T^{\mu}_{(M)\alpha}) - \frac{1}{2}\sqrt{-g}\frac{\partial g_{\mu\nu}}{\partial x^{\alpha}}T^{\mu\nu}_{(M)} = 0 \tag{9}$$

Owing to $\dfrac{\delta(\sqrt{-g}L_G)}{\delta g_{\mu\nu}} = -\dfrac{1}{2}\sqrt{-g}T^{\mu\nu}_{(M)}$, we obtain

$$-\frac{1}{2}\sqrt{-g}\frac{\partial g_{\mu\nu}}{\partial x^{\alpha}}T^{\mu\nu}_{(M)} = \frac{\partial g_{\mu\nu}}{\partial x^{\alpha}}\frac{\delta(\sqrt{-g}L_G)}{\delta g_{\mu\nu}}$$

$$= \frac{\partial g_{\mu\nu}}{\partial x^{\alpha}}\left(\frac{\partial(\sqrt{-g}L_G)}{\partial g_{\mu\nu}} - \frac{\partial}{\partial x^{\lambda}}\frac{\partial(\sqrt{-g}L_G)}{\partial g_{\mu\nu,\lambda}} + \frac{\partial^2}{\partial x^{\lambda}\partial x^{\sigma}}\frac{\partial(\sqrt{-g}L_G)}{\partial g_{\mu\nu,\lambda\sigma}}\right), \tag{10}$$

here we have adopted $\sqrt{-g(x)}L_G(x) = \sqrt{-g(x)}L_G[g_{\mu\nu}(x); g_{\mu\nu,\lambda}(x); g_{\mu\nu,\lambda\sigma}(x)]$ as the Lagrangian density of gravitational field. After some calculations and using the definition Eq.(8), Eq.(10) can be transformed into

$$-\frac{1}{2}\sqrt{-g}\frac{\partial g_{\mu\nu}}{\partial x^{\alpha}}T^{\mu\nu}_{(M)} = \frac{\partial}{\partial x^{\lambda}}\left(\sqrt{-g}L_G\delta^{\lambda}_{\alpha} - \frac{\partial(\sqrt{-g}L_G)}{\partial g_{\mu\nu,\lambda}}g_{\mu\nu,\alpha} - \frac{\partial(\sqrt{-g}L_G)}{\partial g_{\mu\nu,\lambda\sigma}}g_{\mu\nu,\sigma\alpha}\right.$$

$$\left. +\frac{\partial}{\partial x^{\sigma}}(\frac{\partial(\sqrt{-g}L_G)}{\partial g_{\mu\nu,\lambda\sigma}})g_{\mu\nu,\alpha}\right) = \frac{\partial}{\partial x^{\lambda}}(\sqrt{-g}\tilde{t}^{\lambda}_{(G)\alpha}) \tag{11}$$

From Eq.(9) and Eq.(11) Einstein conservation laws $\dfrac{\partial}{\partial x^{\mu}}(\sqrt{-g}T^{\mu}_{(M)\alpha} + \sqrt{-g}\tilde{t}^{\mu}_{(G)\alpha}) = 0$ are obtained immediately.

**B. The method used in the text book "The classical theory of fields" [4] written by Landau and Lifshitz**

The starting point of this method is Bianchi identities and Einstein field equations *i.e.* $G^{\mu\nu} = (R^{\mu\nu} - \frac{1}{2}g^{\mu\nu}R) = -8\pi G T^{\mu\nu}_{(M)}$. If we choose a geodesic coordinate system[7], then Eq.(9) will reduce to $\frac{\partial}{\partial x^{\mu}}(T^{\mu}_{(M)\alpha})=0$, or in contravariant components $\frac{\partial}{\partial x^{\mu}}(T^{\alpha\mu}_{(M)})=0$. In the geodesic coordinate system one can obtain

$$T^{\alpha\beta}=-\frac{1}{16\pi G}\frac{\partial}{\partial x^{\gamma}}\{\frac{1}{(-g)}\frac{\partial}{\partial x^{\delta}}[(-g)(g^{\alpha\beta}g^{\gamma\delta}-g^{\alpha\gamma}g^{\beta\delta})]\}$$
$$=-\frac{1}{16\pi G}\frac{\partial}{\partial x^{\gamma}}\{\frac{1}{\sqrt{-g}}\frac{\partial}{\partial x^{\delta}}[\sqrt{-g}(g^{\alpha\beta}g^{\gamma\delta}-g^{\alpha\gamma}g^{\beta\delta})]\}$$

(12)

Introducing the quantities[4]

$$h^{\alpha\beta\gamma}=-\frac{1}{16\pi G}\frac{\partial}{\partial x^{\delta}}[(-g)(g^{\alpha\beta}g^{\gamma\delta}-g^{\alpha\gamma}g^{\beta\delta})]$$

(13)

which are skew-symmetric in their indices β and γ; $T^{\alpha\mu}_{(M)}$ satisfying $\frac{\partial}{\partial x^{\mu}}(T^{\alpha\mu}_{(M)})=0$ can be written in the form[4] $\frac{\partial h^{\alpha\beta\gamma}}{\partial x^{\gamma}}=(-g)T^{\alpha\beta}_{(M)}$. If we choose an arbitrary coordinate system, $\frac{\partial h^{\alpha\beta\gamma}}{\partial x^{\gamma}}\neq(-g)T^{\alpha\beta}_{(M)}$; the difference between the left-hand side and the right-hand side may be denoted by $(-g)t^{*\alpha\beta}_{(G)}$:

$$(-g)t^{*\alpha\beta}_{(G)}=\frac{\partial h^{\alpha\beta\gamma}}{\partial x^{\gamma}}-(-g)T^{\alpha\beta}_{(M)}$$

(14)

From Eq.(14) we obtain conservation laws

$$\frac{\partial}{\partial x^{\beta}}[(-g)(T^{\alpha\beta}_{(M)}+t^{*\alpha\beta}_{(G)})]=0$$

(15)

immediately; $t^{*\alpha\beta}_{(G)}$ is interpreted as the energy-momentum pseudo tensor of the gravitational field [4].

It must be noted that $\frac{\partial}{\partial x^{\beta}}[(-g)(T^{\alpha\beta}_{(M)}+t^{*\alpha\beta}_{(G)})]=0$ and $\frac{\partial}{\partial x^{\beta}}[\sqrt{-g}(T^{\alpha\beta}_{(M)}+t^{\sim\alpha\beta}_{(G)})]=0$ are not like in form. But by using the same method B we can obtain other conservation laws. Introducing the quantities

$$h^{\#\alpha\beta\gamma}=-\frac{1}{16\pi G}\frac{\partial}{\partial x^{\delta}}[\sqrt{-g}(g^{\alpha\beta}g^{\gamma\delta}-g^{\alpha\gamma}g^{\beta\delta})]$$

(16)

which are also skew-symmetric in their indices β and γ. $h^{\#\alpha\beta\gamma}$ is connected with $h^{\alpha\beta\gamma}$ by the relation

$$h^{\alpha\beta\gamma} = -\frac{1}{16\pi G}(\frac{\partial\sqrt{-g}}{\partial x^{\delta}})\sqrt{-g}(g^{\alpha\beta}g^{\gamma\delta} - g^{\alpha\gamma}g^{\beta\delta}) + \sqrt{-g}h^{\#\alpha\beta\gamma} \tag{17}$$

From Eqs. (12, 16) we note that $T^{\alpha\mu}_{(M)}$ satisfying $\frac{\partial}{\partial x^{\mu}}(T^{\alpha\mu}_{(M)}) = 0$ in the geodesic coordinate system can be also written in the form $\frac{\partial h^{\#\alpha\beta\gamma}}{\partial x^{\gamma}} = \sqrt{-g}T^{\alpha\beta}_{(M)}$. If we choose an arbitrary coordinate system, $\frac{\partial h^{\#\alpha\beta\gamma}}{\partial x^{\gamma}} \neq \sqrt{-g}T^{\alpha\beta}_{(M)}$; the difference between the left-hand side and the right-hand side may be denoted by $\sqrt{-g}t^{\#\alpha\beta}_{(G)}$:

$$\sqrt{-g}t^{\#\alpha\beta}_{(G)} = \frac{\partial h^{\#\alpha\beta\gamma}}{\partial x^{\gamma}} - \sqrt{-g}T^{\alpha\beta}_{(M)} \tag{18}$$

From Eq.(18) we **immediately** obtain conservation laws

$$\frac{\partial}{\partial x^{\beta}}[\sqrt{-g}(T^{\alpha\beta}_{(M)} + t^{\#\alpha\beta}_{(G)})] = 0 \tag{19}$$

**these** conservation laws **are** like in form with $\frac{\partial}{\partial x^{\beta}}[\sqrt{-g}(T^{\alpha\beta}_{(M)} + t^{\sim\alpha\beta}_{(G)})] = 0$.

From Eqs. (13, 16) and Eqs. (14,17), after some calculations, it is not difficult to get

$$\sqrt{-g}(t^{\#\alpha\beta}_{(G)} - t^{*\alpha\beta}_{(G)}) = \frac{\partial h^{\#\alpha\beta\gamma}}{\partial x^{\gamma}} - \frac{1}{\sqrt{-g}}\frac{\partial h^{\alpha\beta\gamma}}{\partial x^{\gamma}}$$

$$= \frac{\partial h^{\#\alpha\beta\gamma}}{\partial x^{\gamma}} + \frac{1}{\sqrt{-g}}\frac{\partial}{\partial x^{\gamma}}\{\frac{1}{16\pi G}[\frac{\partial(\sqrt{-g})}{\partial x^{\delta}}]\sqrt{-g}(g^{\alpha\beta}g^{\gamma\delta} - g^{\alpha\gamma}g^{\beta\delta})\}$$

$$-\frac{1}{\sqrt{-g}}\frac{\partial}{\partial x^{\gamma}}(\sqrt{-g}h^{\#\alpha\beta\gamma}) \tag{20}$$

$$= \frac{1}{\sqrt{-g}}\frac{\partial}{\partial x^{\gamma}}\{\frac{1}{16\pi G}[\frac{\partial(\sqrt{-g})}{\partial x^{\delta}}]\sqrt{-g}(g^{\alpha\beta}g^{\gamma\delta} - g^{\alpha\gamma}g^{\beta\delta})\} - \frac{1}{\sqrt{-g}}\frac{\partial(\sqrt{-g})}{\partial x^{\gamma}}h^{\#\alpha\beta\gamma}$$

Eq.(15) and Eq.(19) are Einstein conservation laws derived by the method of Landau and Lifshitz; and Eq.(15) is related to Eq.(19) through Eq.(17) and Eq.(20); hence these two conservation laws are equivalent.

Since $\frac{\partial}{\partial x^\beta}[(-g)(T^{\alpha\beta}_{(M)} + t^{*\alpha\beta}_{(G)})] = 0$ and $\frac{\partial}{\partial x^\beta}[\sqrt{-g}(T^{\alpha\beta}_{(M)} + t^{\sim\alpha\beta}_{(G)})] = 0$ are not like in form, but

$\frac{\partial}{\partial x^\beta}[\sqrt{-g}(T^{\alpha\beta}_{(M)} + t^{\#\alpha\beta}_{(G)})] = 0$ and $\frac{\partial}{\partial x^\beta}[\sqrt{-g}(T^{\alpha\beta}_{(M)} + t^{\sim\alpha\beta}_{(G)})] = 0$ are like in form; we shall only discuss $\frac{\partial}{\partial x^\beta}[\sqrt{-g}(T^{\alpha\beta}_{(M)} + t^{\#\alpha\beta}_{(G)})] = 0$ in this paper. $\sqrt{-g}t^{\#\alpha\beta}_{(G)}$ and $\sqrt{-g}t^{\sim\alpha\beta}_{(G)}$ are connected by the relation $\sqrt{-g}t^{\#\alpha\beta}_{(G)} = \sqrt{-g}t^{\sim\alpha\beta}_{(G)} + \frac{\partial}{\partial x^\lambda}w^{\#\alpha\beta\lambda}_{(G)}$, where

$\frac{\partial}{\partial x^\lambda}w^{\#\alpha\beta\lambda}_{(G)} = -\frac{\partial}{\partial x^\lambda}w^{\#\alpha\lambda\beta}_{(G)}$; $\sqrt{-g}t^{\#\alpha\beta}_{(G)}$ and $\sqrt{-g}t^{\sim\alpha\beta}_{(G)}$ have the relation of equivalence class[2,8].

**C. The method of Weinberg**

Weinberg had derived a conservation law in his book "Gravitation and Cosmology" [5], but the procession of his derivation is incomplete and not exact because he made the left hand side of the used field equations linearization but not made its right hand side linearization. In this paper, we shall make some changes in the derivation. As it is basically modeled on the derivation of Weinberg, so we shall still call it the method of Weinberg.

The starting point of this method is also Bianchi identities

$$(\sqrt{-g}G^\mu_\alpha)_{;\mu} = \frac{\partial}{\partial x^\mu}(\sqrt{-g}G^\mu_\alpha) - \frac{1}{2}\sqrt{-g}\frac{\partial g_{\mu\nu}}{\partial x^\alpha}G^{\mu\nu} = 0 \tag{21}$$

and Einstein field equations: $G^{\mu\nu} = (R^{\mu\nu} - \frac{1}{2}g^{\mu\nu}R) = -8\pi G T^{\mu\nu}_{(M)}$. If we choose a geodesic coordinate system, then Eq. (21) will reduce to $\frac{\partial}{\partial x^\mu}(G^\mu_\alpha) = 0$, or in contravariant components $\frac{\partial}{\partial x^\mu}(G^{\alpha\mu}) = 0$.

In the geodesic coordinate system one can obtain

$$G^{\alpha\beta} = \frac{1}{2}\frac{\partial}{\partial x^\gamma}\{\frac{1}{(-g)}\frac{\partial}{\partial x^\delta}[(-g)(g^{\alpha\beta}g^{\gamma\delta} - g^{\alpha\gamma}g^{\beta\delta})]\}$$
$$= \frac{1}{2}\frac{\partial}{\partial x^\gamma}\{\frac{1}{\sqrt{-g}}\frac{\partial}{\partial x^\delta}[\sqrt{-g}(g^{\alpha\beta}g^{\gamma\delta} - g^{\alpha\gamma}g^{\beta\delta})]\} \tag{22}$$

Introducing the quantities

$$H^{\alpha\beta\gamma}=\frac{1}{2}\frac{\partial}{\partial x^\delta}[\sqrt{-g}(g^{\alpha\beta}g^{\gamma\delta}-g^{\alpha\gamma}g^{\beta\delta})] \tag{23}$$

which are skew-symmetric in their indices β and γ; $G^{\alpha\mu}$ satisfying $\frac{\partial}{\partial x^\mu}(G^{\alpha\mu})=0$ can be written in the form $\frac{\partial H^{\alpha\beta\gamma}}{\partial x^\gamma}=\sqrt{-g}G^{\alpha\beta}$. Let $\sqrt{-g}G^{(1)\alpha\beta}=\frac{\partial H^{\alpha\beta\gamma}}{\partial x^\gamma}$. If we choose an arbitrary coordinate system, $\frac{\partial h^{\alpha\beta\gamma}}{\partial x^\gamma}\neq\sqrt{-g}G^{\alpha\beta}$. Define

$$\sqrt{-g}t^{\wedge\alpha\beta}_{(G)}=\frac{\sqrt{-g}}{8\pi G}(G^{\alpha\beta}-G^{(1)\alpha\beta}) \tag{24}$$

Then the Einstein field equations $G^{\alpha\beta}=-8\pi G T^{\alpha\beta}_{(M)}$ can be rewritten as

$$\sqrt{-g}(G^{\alpha\beta}-8\pi G t^{\wedge\alpha\beta}_{(G)})=-8\pi G\sqrt{-g}(T^{\alpha\beta}_{(M)}+t^{\wedge\alpha\beta}_{(G)}) \tag{25}$$

Owing to $\sqrt{-g}(G^{\alpha\beta}-8\pi G t^{\wedge\alpha\beta}_{(G)})=\sqrt{-g}G^{(1)\alpha\beta}=\frac{\partial H^{\alpha\beta\gamma}}{\partial x^\gamma}$, from Eq. (25) we obtain conservation laws

$$\frac{\partial}{\partial x^\beta}[\sqrt{-g}(T^{\alpha\beta}_{(M)}+t^{\wedge\alpha\beta}_{(G)})]=0 \tag{26}$$

immediately; $t^{\wedge\alpha\beta}_{(G)}$ is interpreted as the energy-momentum pseudo tensor of the gravitational field[5]. We see that $\frac{\partial}{\partial x^\beta}[\sqrt{-g}(T^{\alpha\beta}_{(M)}+t^{\wedge\alpha\beta}_{(G)})]=0$ and $\frac{\partial}{\partial x^\beta}[\sqrt{-g}(T^{\alpha\beta}_{(M)}+t^{\sim\alpha\beta}_{(G)})]=0$ are like in form; $\sqrt{-g}t^{\wedge\alpha\beta}_{(G)}$ and $\sqrt{-g}t^{\sim\alpha\beta}_{(G)}$ are connected by the relation

$$\sqrt{-g}t^{\wedge\alpha\beta}_{(G)}=\sqrt{-g}t^{\sim\alpha\beta}_{(G)}+\frac{\partial}{\partial x^\lambda}w^{\wedge\alpha\beta\lambda}_{(G)}\text{, where}$$

$$\frac{\partial}{\partial x^\lambda} w_{(G)}^{\wedge \alpha\beta\lambda} = -\frac{\partial}{\partial x^\lambda} w_{(G)}^{\wedge \alpha\lambda\beta} \; ; \; \sqrt{-g}\, t_{(G)}^{\wedge \alpha\beta} \text{ and } \sqrt{-g}\, t_{(G)}^{\sim \alpha\beta}$$ have the relation of equivalence class [2,8].

On the surface, the above methods for the derivation of Einstein conservation laws and the relevant definitions of energy-momentum tensor density for gravitational field are dissimilar from each other, but in essence they are all equivalent.

### 3. The influences of conservation laws upon the characteristics of the gravitational waves

In this section we shall study the influences both of Lorentz and Levi-Civita conservation laws and Einstein conservation laws upon the characteristics of the gravitational waves.

### 3.1 If Lorentz and Levi-Civita conservation laws are correct

The relation $\sqrt{-g}\, T^\mu_{(M)\alpha}(x) + \sqrt{-g}\, T^\mu_{(G)\alpha}(x) = 0$ must exist ; hence

$\sqrt{-g}\, T^0_{(M)0}(x) + \sqrt{-g}\, T^0_{(G)0}(x) = 0$ , $\sqrt{-g}\, T^i_{(M)0}(x) + \sqrt{-g}\, T^i_{(G)0}(x) = 0$ . From Eq.(1) we get:

$$-\frac{\partial}{\partial t}\int_V (\sqrt{-g}\, T^0_{(M)0}(x) + \sqrt{-g}\, T^0_{(G)0}(x))dV$$
$$= C\oiint_S (\sqrt{-g}\, T^i_{(M)0}(x) + \sqrt{-g}\, T^i_{(G)0}(x))dS_i \quad (27)$$

The left hand side of Eq. (27) represents the variation rate of the total energy in the volume V; the right hand side of Eq. (27) represents the total momentum current across the closed surface S which surrounds the volume V. Evidently

$$-\frac{\partial}{\partial t}\int_V (\sqrt{-g}\, T^0_{(M)0}(x) + \sqrt{-g}\, T^0_{(G)0}(x))dV = 0 , \quad (28)$$

and $\quad C\oiint_S (\sqrt{-g}\, T^i_{(M)0}(x) + \sqrt{-g}\, T^i_{(G)0}(x))dS_i = 0 \quad (29)$

These two relations indicate that the total energy in the volume V does not change and the total momentum current across the closed surface S is always zero.

### 3.2 If Einstein conservation laws are correct

As we have shown above, there are three different forms for Einstein conservation laws:

$$\frac{\partial}{\partial x^\beta}[\sqrt{-g}(T^{\alpha\beta}_{(M)} + t^{\sim\alpha\beta}_{(G)})] = 0, \quad \frac{\partial}{\partial x^\beta}[\sqrt{-g}(T^{\alpha\beta}_{(M)} + t^{\#\alpha\beta}_{(G)})] = 0 \text{ and}$$

$$\frac{\partial}{\partial x^\beta}[\sqrt{-g}(T^{\alpha\beta}_{(M)} + t^{\wedge\alpha\beta}_{(G)})] = 0 .$$ Since they are all equivalent in essence, we will treat only

$\frac{\partial}{\partial x^\beta}[\sqrt{-g}(T^{\alpha\beta}_{(M)} + t^{\sim\alpha\beta}_{(G)})] = 0$ in the following.

Since there are the relations

$$\sqrt{-g}T^\mu_{(M)\alpha} + \sqrt{-g}t^{\sim\mu}_{(G)\alpha} = \sqrt{-g}T^\mu_{(M)\alpha} + \sqrt{-g}T^\mu_{(G)\alpha} - \frac{\partial}{\partial x^\beta}u^{\mu\beta}_{(G)\alpha}$$

$$= -\frac{\partial}{\partial x^\beta}u^{\mu\beta}_{(G)\alpha} \left(= \frac{\partial}{\partial x^\beta}u^{\beta\mu}_{(G)\alpha}\right)$$

we have

$$\sqrt{-g}T^0_{(M)0} + \sqrt{-g}t^{\sim 0}_{(G)0} = -\frac{\partial}{\partial x^\beta}u^{0\beta}_{(G)0}, \quad \sqrt{-g}T^i_{(M)0} + \sqrt{-g}t^{\sim i}_{(G)0} = -\frac{\partial}{\partial x^\beta}u^{i\beta}_{(G)0}.$$

From Eq.(3) we get:

$$-\frac{\partial}{\partial t}\int_V (\sqrt{-g}T^0_{(M)0}(x) + \sqrt{-g}t^{\square 0}_{(G)0}(x))dV$$
$$= C\oiint_S (\sqrt{-g}T^i_{(M)0}(x) + \sqrt{-g}t^{\square i}_{(G)0}(x))dS_i \tag{30}$$

Owing to $\frac{\partial}{\partial x^\beta}u^{\mu\beta}_{(G)\alpha} = -\frac{\partial}{\partial x^\beta}u^{\beta\mu}_{(G)\alpha}$, the following relations can be derived:

$$-\frac{\partial}{\partial t}\int_V (\sqrt{-g}T^0_{(M)0}(x) + \sqrt{-g}t^{\square 0}_{(G)0}(x))dV = 0 \tag{31}$$

$$C\oiint_S (\sqrt{-g}T^i_{(M)0}(x) + \sqrt{-g}t^{\square i}_{(G)0}(x))dS_i = 0 \tag{32}$$

These two relations also indicate that, for Einstein conservation laws, the total energy in the volume V does not change and the total momentum current across the closed surface S is still always zero.

**3.3 The changes and the transformations of energy in a gravitational system when gravitational waves exist**

From the above analysis for Lorentz conservation laws and Einstein conservation laws, we see that no matter which conservation laws are correct, the gravitational energy and momentum do not spread in a gravitational system. But gravitational waves should exist, this is because: according to Einstein field equations

$\sqrt{-g}(R^{\mu\nu} - \frac{1}{2}g^{\mu\nu}R) = -8\pi G\sqrt{-g}T^{\mu\nu}_{(M)}$, when the energy-momentum tensor density $\sqrt{-g}T^{\mu\nu}_{(M)}$ of the wave source is changed for some causes (such as supernova explosions, orbital change of binary pulsar, etc.), the curvature of space-time in the neighboring regions of the source would change subsequently, the essential cause is the change of metric tensor field $g_{\mu\nu}(x)$. Since the characteristic equation of Einstein field equations is the same equation as the characteristic equation of D'Alembert equation [9], and the characteristic equation of D'Alembert equation indicate the law of propagation for wave front; therefore the changes of the metric tensor field $g_{\mu\nu}(x)$ and the curvature of space-time must also propagate with the speed of light from the wave source to far distance. Then the gravitational waves are created.

We must point out that although the gravitational waves do not bring energy and momentum spread, however within the gravitational system there are still energy changes and energy transformations. As an example to interpret this problem, below we shall discuss the energy changes and energy transformations of a Weber cylinder when it receives gravitational waves.

There are different kinds of matter field energy for the Weber cylinder, they include the gravitational potential energy of this cylinder $\overset{1}{T}{}^0_{(M)0}$ ($\overset{1}{T}{}^0_{(M)0}$ should be one kind of matter field energy [10]), the vibration energy of this cylinder $\overset{2}{T}{}^0_{(M)0}$, and the thermal energy of this cylinder $\overset{3}{T}{}^0_{(M)0}$, etc. Then we have $T^0_{(M)0} = \overset{1}{T}{}^0_{(M)0} + \overset{2}{T}{}^0_{(M)0} + \overset{3}{T}{}^0_{(M)0} + \cdots$. When gravitational waves come, metric tensor field and the curvature of space-time in the neighboring regions of the Weber cylinder will change; therefore the energy of free gravitational field $T^0_{(G)0}$ or $\tilde{t}^0_{(G)0}$ in the neighboring regions of this cylinder will change subsequently. From Eq. (28) or Eq. (31), we obtain:

$$-\frac{\partial}{\partial t}\int_V (\sqrt{-g}\,\overset{1}{T}{}^0_{(M)0}(x) + \sqrt{-g}\,\overset{2}{T}{}^0_{(M)0}(x) + \sqrt{-g}\,\overset{3}{T}{}^0_{(M)0}(x) + \cdots + \sqrt{-g}\,T^0_{(G)0}(x))dV = 0 \quad (33)$$

or $\quad -\frac{\partial}{\partial t}\int_V (\sqrt{-g}\,\overset{1}{T}{}^0_{(M)0}(x) + \sqrt{-g}\,\overset{2}{T}{}^0_{(M)0}(x) + \sqrt{-g}\,\overset{3}{T}{}^0_{(M)0}(x) + \cdots + \sqrt{-g}\,\tilde{t}^0_{(G)0}(x))dV = 0 \quad (34)$

where V is the volume of the Weber cylinder. Eq. (33) and Eq. (34) tell us that when gravitational waves come, the matter field energy of the Weber cylinder $T^0_{(M)0}$ will also change. Eq. (33) can be transformed into

$$\frac{\partial}{\partial t}\int_V (\sqrt{-g}\,\overset{2}{T}{}^0_{(M)0}(x))dV = -\frac{\partial}{\partial t}\int_V (\sqrt{-g}\,\overset{1}{T}{}^0_{(M)0}(x) + \sqrt{-g}\,\overset{3}{T}{}^0_{(M)0}(x) + \cdots + \sqrt{-g}\,T^0_{(G)0}(x))dV \quad (35)$$

and Eq. (34) can be transformed into

$$\frac{\partial}{\partial t}\int_V (\sqrt{-g}\,\overset{2}{T}{}^0_{(M)0}(x))dV = -\frac{\partial}{\partial t}\int_V (\sqrt{-g}\,\overset{1}{T}{}^0_{(M)0}(x) + \sqrt{-g}\,\overset{3}{T}{}^0_{(M)0}(x) + \cdots + \sqrt{-g}\,\tilde{t}^0_{(G)0}(x))dV \quad (36)$$

Eq.(35) (or Eq.(36)) shows that the increase of the vibration energy of Weber cylinder accounts for the decreases of the gravitational potential energy of Weber cylinder and the thermal energy of Weber cylinder and the energy of free gravitational field $T^0_{(G)0}$ (or $\tilde{t}^{\,0}_{(G)0}$ ).

**4. A new explanation about the orbital decay of PSR1913+16**

Many people think that the observation for the orbital period decay rate of PSR1913+16 verifies that the gravitational waves might carrying energy and momentum in their propagation; the author had shown that these views are not accurate in Ref.[11,12,13]. We have explained in the above section that gravitational waves do not bring energy and momentum spread. Then how can we give an explanation about the orbital period decay of PSR1913+16? We shall give a new explanation in this section.

PSR1913+16 is a binary pulsar. The matter field energy of PSR1913+16 $T^0_{(M)0}$ includes the gravitational potential energy of this binary pulsar $^A T^0_{(M)0}$ ($^A T^0_{(M)0}$ should be one kind of matter field energy), the orbital kinetic energy of this binary pulsar $^B T^0_{(M)0}$, the thermal energy of this binary pulsar $^C T^0_{(M)0}$ ,etc. then we have $T^0_{(M)0} = {^A T^0_{(M)0}} + {^B T^0_{(M)0}} + {^C T^0_{(M)0}} + \cdots$ .From classical mechanics we can obtain

$$\frac{\partial}{\partial t}\int_V (\sqrt{-g}\,{^A T^0_{(M)0}}(x) + \sqrt{-g}\,{^B T^0_{(M)0}}(x))dV = \frac{d}{dt}(-\frac{Gm_1m_2}{R} + \frac{1}{2}\frac{m_1m_2}{m_1+m_2}\omega^2 R^2) \qquad (37)$$

Where $m_1, m_2$ are the masses for the two stars of PSR1913+16, R is their distance, $P = 2\pi/\omega$ is the orbital period. Observation results show that $\frac{d}{dt}(-\frac{Gm_1m_2}{R} + \frac{1}{2}\frac{m_1m_2}{m_1+m_2}\omega^2 R^2) < 0$, therefore the gravitational potential energy and the orbital kinetic energy of PSR1913+16 will decrease. What is the cause of these energy decreases? In the following we shall use Lorentz and Levi-Civita conservation laws and Einstein conservation laws to explain this problem.

If we use Lorentz and Levi-Civita conservation laws from Eq. (28) there must have

$$\frac{\partial}{\partial t}\int_V (\sqrt{-g}\,{^A T^0_{(M)0}}(x) + \sqrt{-g}\,{^B T^0_{(M)0}}(x) + \sqrt{-g}\,{^C T^0_{(M)0}}(x) + \cdots + \sqrt{-g}\,T^0_{(G)0}(x))dV = 0 \qquad (38)$$

Putting Eq. (37) into Eq. (38), we immediately get

$$\frac{\partial}{\partial t}\int_V (\sqrt{-g}\overset{C}{T}^0_{(M)0}(x)+\cdots+\sqrt{-g}T^0_{(G)0}(x))dV = -\frac{\partial}{\partial t}\int_V (\sqrt{-g}\overset{A}{T}^0_{(M)0}(x)+\sqrt{-g}\overset{B}{T}^0_{(M)0}(x))dV \qquad (39)$$

$$= -\frac{d}{dt}(-\frac{Gm_1 m_2}{R}+\frac{1}{2}\frac{m_1 m_2}{m_1+m_2}\omega^2 R^2)$$

If we use Einstein conservation laws there must have similarly

$$\frac{\partial}{\partial t}\int_V (\sqrt{-g}\overset{A}{T}^0_{(M)0}(x)+\sqrt{-g}\overset{B}{T}^0_{(M)0}(x)+\sqrt{-g}\overset{C}{T}^0_{(M)0}(x)+\cdots+\sqrt{-g}\tilde{t}^0_{(G)0}(x))dV = 0 \qquad (40)$$

Putting Eq. (37) into Eq. (40), we immediately get

$$\frac{\partial}{\partial t}\int_V (\sqrt{-g}\overset{C}{T}^0_{(M)0}(x)+\cdots+\sqrt{-g}\tilde{t}^0_{(G)0}(x))dV = -\frac{\partial}{\partial t}\int_V (\sqrt{-g}\overset{A}{T}^0_{(M)0}(x)+\sqrt{-g}\overset{B}{T}^0_{(M)0}(x))dV \qquad (41)$$

$$= -\frac{d}{dt}(-\frac{Gm_1 m_2}{R}+\frac{1}{2}\frac{m_1 m_2}{m_1+m_2}\omega^2 R^2)$$

Eq.(39) or Eq.(41) tell us some gravitational potential energy and orbital kinetic energy of PSR1913+16 are transformed into the thermal energy of this binary pulsar and the energy of free gravitational field ($T^0_{(G)0}$ or $\tilde{t}^0_{(G)0}$), so the gravitational potential energy and the orbital kinetic energy of PSR1913+16 will decrease.

## 5. The characteristics of Lorentz and Levi-Civita conservation laws differ from that of Einstein conservation laws

In sections 3 and 4 we talked about the identical characteristics of the Lorenz and Levi-Civita conservation laws and the Einstein conservation laws. But there are different characteristics in these two conservation laws: Lorenz and Levi-Civita conservation laws affirm that the sum of the energy-momentum tensor density for the matter field and the energy-momentum tensor density for the gravitational field are equal to zero, *i.e.* $\sqrt{-g}T^\mu_{(M)\alpha}+\sqrt{-g}T^\mu_{(G)\alpha}=0$; Einstein conservation laws affirm that the sum of the energy-momentum tensor density for the matter field and the energy-momentum tensor density for the gravitational field are not equal to zero, and are

$$\sqrt{-g}T^\mu_{(M)\alpha}+\sqrt{-g}\tilde{t}^\mu_{(G)\alpha}=-\frac{\partial}{\partial x^\beta}u^{\mu\beta}_{(G)\alpha}\,(=\frac{\partial}{\partial x^\beta}u^{\beta\mu}_{(G)\alpha})\;\text{constantly.}$$

When Lorenz and Levi-Civita conservation laws are correct, under certain conditions, some positive energy of matter field and some negative energy of gravitational field might annihilate mutually to zero energy. This possibility could be used to establish the speculation that the event horizon might not appear hence the black hole could not form [14]. Under other conditions, some positive energy of matter field and some negative energy of gravitational field might be produced from zero energy. This possibility could be used to explain the hypotheses that consider that the 'big bang' might not exist and consider that the positive energy of matter field might be created with the negative energy of gravitational field from zero energy [1, 15]. These speculation and hypotheses will lead to some observable phenomena.

When Einstein conservation laws are correct, the above speculation and hypotheses and the observable phenomena led by them do not exist. So from whether or not these observable phenomena exist, we can judge which conservation laws are correct.

As to which conservation laws are correct must be ultimately determined by observable and experimental results; but based on theoretical analysis, we can also make some deductions.

$\sqrt{-g}\tilde{t}^{\mu}_{(G)\alpha}$ is a pseudo tensor density, it and the Einstein conservation laws $\frac{\partial}{\partial x^{\beta}}[\sqrt{-g}(T^{\alpha\beta}_{(M)}+\tilde{t}^{\alpha\beta}_{(G)})]=0$ all lack the invariant character which should have in the spirit of general relativity. In addition, $\sqrt{-g}\tilde{t}^{\mu}_{(G)\alpha}$ is a multivalued quantity; $\sqrt{-g}\tilde{t}^{\mu}_{(G)\alpha}$ or $\sqrt{-g}\tilde{t}^{\lambda}_{(G)\alpha}-\frac{\partial}{\partial x^{\mu}}V^{\lambda\mu}_{\alpha}$ ( $V^{\lambda\mu}_{\alpha}$ are any function which satisfy $\frac{\partial}{\partial x^{\mu}}V^{\lambda\mu}_{\alpha}=-\frac{\partial}{\partial x^{\mu}}V^{\mu\lambda}_{\alpha}$ ) all can be regarded as energy-momentum pseudo tensor density for the gravitational field. But $\sqrt{-g}T^{\mu}_{(M)\alpha}$ is a tensor density and is a single valued quantity, it and the Lorenz and Levi-Civita conservation laws *i.e.* $\frac{\partial}{\partial x^{\mu}}(\sqrt{-g}T^{\mu}_{(M)\alpha}+\sqrt{-g}T^{\mu}_{(G)\alpha})=0$ and $\sqrt{-g}T^{\mu}_{(M)\alpha}+\sqrt{-g}T^{\mu}_{(G)\alpha}=0$ all have the invariant character. Obviously the Lorenz and Levi-Civita conservation laws are better than the Einstein conservation laws on scientific logic.

**References**

[1] Chen F. P. 2008, "Field equations and conservation laws derived from the generalized Einstein's Lagrangian density for a gravitational system and their implications to cosmology." Int.J.Theor.Phys.47, 421.
[2] Chen F. P. 2008, "The covariant energy-momentum tensor densities and its conservation laws for a gravitational system I ." *Science paper online* 200802-56(in Chinese).
[3] Chen F. P. 2008, "A further generalized Lagrangian density and its special cases."

*Int.J.Theor.Phys.* published online first.

[4]. Landau L. and Lifshitz E. 1975, "The Classical Theory of Fields", Translated by M. Hamermesh, Pergamon Press, Oxford.

[5]. Weinberg S. 1972, "Gravitation and Cosmology", Wiley, New York.

[6] Moller C. 1972, "The Theory of Relativity", Clarendon Press, Oxford.

[7] Carmeli M. 1982,. "Classical Fields: General Relativity and Gauge Theory", John Wiley. & Sons, New York.

[8] Chen F. P. 2002, "The restudy on the debate between Einstein and Levi-Civita and the experimental tests." Spacetime & Substance. 3, 161.

[9] Fock V. 1959, "The theory of space, time, and gravitation", Pergamon Press, London.

[10] Chen F. P. 2002, "The restudy on the debate between Einstein and Levi-Civita and the experimental tests." Spacetime & Substance. 3, 161.

[11]. Chen F. P. 1997, "PSR1913+16 does not verify the gravitational radiation", Ziran Zazhi (Journal of Nature),19, 243 (in Chinese ).

[12] Chen F. P. 1998, "Investigation on problem of coincidence between the observed and the predicted value of orbital period decay rate $\dot{P}_b$ for PSR1534+12", Ziran Zazhi (Journal of Nature), 20,178 (in Chinese ).

[13] Chen F. P. 2003, "A new study about the two formulations of conservation laws for matter plus gravitational field and their experimental test", arXiv, gr-qc/0003008.

[14] Chen F.P. 2008, "The Lorentz and Lavi-Civita conservation laws prohibit the existence of black holes", arXiv,08052451.

[15] Chen F. P. 2006, "A new theory of cosmology that preserves the generally recognized symmetries of cosmos, explains the origin of the energy for the matter field, but excludes the existence of Big Bang."arXiv, gr-qc/0605076.